# Uncertainty prediction of built-up areas from global human settlement data in the United States based on landscape metrics

Johannes H. Uhl[1,2], Stefan Leyk[2,3]

[1]University of Colorado Boulder, Cooperative Institute for Research in Environmental Sciences (CIRES),
216 UCB, Boulder, CO-80309, USA.
[2]University of Colorado Boulder, Institute of Behavioral Science, 483 UCB, Boulder, CO-80309, USA.
[3]University of Colorado Boulder, Department of Geography, 260 UCB, Boulder, CO-80309, USA.
Email: {Johannes.Uhl;Stefan.Leyk}@colorado.edu
Corresponding author: Johannes H. Uhl (Johannes.Uhl@colorado.edu)

**Abstract:** It is common knowledge that the level of landscape heterogeneity may affect the performance of remote sensing based land use / land cover classification. While this issue has been studied in depth for land cover data in general, the specific relationship between the mapping accuracy of built-up surfaces and morphological characteristics of built-up areas has not been analyzed explicitly, an urgent need given the recent emergence of a variety of global, fine-resolution settlement datasets. Moreover, previous studies typically rely on aggregated, broad-scale landscape metrics to quantify the morphology of built-up areas, neglecting the fine-grained spatial variation and scale dependency of such metrics. Herein, we aim to fill this knowledge gap by assessing the associations between localized (focal) landscape metrics, derived from binary built-up surfaces and localized data accuracy estimates. We test our approach for built-up surfaces from the Global Human Settlement Layer (GHSL) for Massachusetts (USA). Specifically, we examine the explanatory power of landscape metrics for predictive modeling of both commission and omission errors in the multi-temporal GHS-BUILT R2018A data product. We find that the Landscape Shape Index (LSI) calculated in focal windows exhibits, in average, the highest levels of correlation to focal accuracy measures. These relationships are scale-dependent, and become stronger with increasing level of spatial support. Our results are highly consistent across different geographic regions within the U.S., and we find that thematic omission error, as measured by Recall, has the strongest relationship to measures of built-up surface morphology across different temporal epochs and spatial resolutions. The results of our regression analysis ($R^2$>0.9) indicate that it is possible to estimate commission errors in the GHSL in the absence of reference data, and that omission errors in the GHSL can be modeled without accessing the data themselves. Lastly, we test the generalizability of our findings by applying our predictive accuracy models to a different version of the GHSL (i.e., the GHS-BUILT-S2) covering a study area in North Carolina. We find varying levels of model transferability that increases with the increasing spatial support at which focal landscape metrics and localized accuracy estimates are calculated.

## 1. Introduction

In order to analyze the dynamics of human settlements on Earth, researchers typically rely on multi-temporal, remote-sensing-derived, gridded built-up surface datasets, such as the Global Human Settlement Layer (GHSL, Pesaresi et al. 2013), the Global Rural-Urban Mapping Project (GRUMP, Balk et al. 2005), the Global Artificial Impervious Area dataset (GAIA, Gong et al. 2020), or the World Settlement Footprint Evolution dataset (Marconcini et al. 2020). In order to develop an unbiased understanding of the human settlement trends measured by these data, thorough knowledge of the uncertainty inherent in these multi-temporal datasets is crucial. However, the quantification of uncertainty, typically done by means of map comparison to an independently compiled reference dataset of presumably higher accuracy, is often impeded by lack of such reference data over large spatial extents (See at al. 2022), in particular for early points in time (Uhl & Leyk 2022a).
Moreover, accuracy estimates of such datasets are often sample-based, and provided as aggregated or region-specific estimates, likely to ignore the fine-scale spatial non-stationarity of data accuracy (Foody 2007, Wickham et al. 2018). This is surprising, as it is well-known that the accuracy of remote-sensing-derived data land use / land cover data products is related to structural landscape characteristics such as the level of landscape segregation or the patch size of urban land (Smith et al. 2002, Smith et al. 2003, Mück et al. 2017). In the same vein, Degen et al. (2018) show that the level of landscape heterogeneity affects the quantization of multispectral remote sensing data such as Landsat data. Furthermore, previous research has shown that the accuracy of built-up surface layers varies regionally (Klotz et al. 2016, Liu et al. 2020), and across the rural-urban continuum (Leyk et al. 2018), which is strongly related to morphological characteristics of landscapes in general (Vizzari 2011, Vizzari et al. 2013) and of settlements in particular (Cyriac 2022).

To account for these spatial variations in accuracy, researchers have started to use spatially explicit accuracy assessments (e.g., Löw et al. 2013, Khatami et al. 2017, Waldner et al. 2017, Mitchell et al. 2018, Morales-Barquero et al. 2019, Uhl & Leyk 2022b) which are based on locally constrained confusion matrices (Foody 2007). Moreover, in order to account for the scarcity of reference data, their resource-intensive creation, and the spatial non-stationarity of geospatial data accuracy, researchers have developed a wide range of methods for predictive accuracy modeling of geospatial data such as land cover data using a variety of techniques and explanatory variables (e.g., Steele et al. 1998, Kyriakidis and Dungan 2001, Smith et al. 2003, Leyk and Zimmermann 2004, van Oort et al. 2004, Comber et al. 2012, Tsutsumida and Comber 2015, Zhang and Mei 2016, Wickham et al. 2018, Mei et al. 2019, Ebrahimy et al. 2021, Cheng et al. 2021), while others have incorporated landscape metrics in land cover data accuracy assessments (Smith et al. 2002 and 2003, Gu & Congalton 2020). Such studies typically focus on land cover data in general, and have not been applied to built-up surface data specifically. Herein, we make use of a multi-temporal reference dataset (i.e., the multi-temporal building footprint dataset for 33 U.S. counties (MTBF-33, Uhl & Leyk 2022a), enabling the creation of historical snapshots of built-up areas at fine spatial and temporal grain, for relatively large, contiguous regions. Using this reference dataset, we conducted a spatially exhaustive, localized accuracy assessment of the Global Human Settlement Layer (GHS-BUILT R2018A, Florczyk et al. 2019) in the state of Massachusetts (USA), for the epochs 1975 and 2014. Consistent to these multi-temporal, continuous surfaces of localized data accuracy estimates, we calculated focal landscape metrics for a large sample of locations (N=200,000 locations) to characterize the morphology of built-up areas. We used these data to (a) assess the association between localized data accuracy and landscape metrics at fine spatial grain, and over time, and (b) test the predictability of localized GHSL data accuracy based on morphological characteristics of both the reference data and the GHSL itself, using two different regression techniques. Finally, we test the sensitivity of our results to the spatial support (i.e., the spatial sample used for focal / localized accuracy and landscape metrics computation) and to the assessment unit (i.e., the spatial resolution of the grid in which accuracy and landscape metrics are computed). Moreover, we analyze the domain adaptation (You et al. 2019) capabilities of our regression models to a different dataset (i.e., the GHS-BUILT-S2, Corbane et al. 2021) and to a study area outside of Massachusetts.

This paper is structured as follows: In Section 2, we discuss the data and methods used, in Section 3, we present and discuss our results, and report our conclusions in Section 4.

## 2. Data and methods

In this section we introduce the used datasets and preprocessing steps undertaken (Section 2.1), as well as the methods used in the different parts of our analyses (Section 2.2).

### 2.1. Data and preprocessing

This study is based on gridded built-up surface layers from the GHSL project and on the multi-temporal building footprint dataset for 33 U.S. counties (MTBF-33, Uhl & Leyk 2022).

*2.1.1. Global Human Settlement Layer (GHS-BUILT)*

The GHS-BUILT R2018A dataset, which is derived from Landsat and Sentinel-2 data, and maps built-up areas at a spatial resolution of 30m, at a global extent, for the years (i.e., epochs) 1975, 1990, 2000, and 2014 (Florczyk et al. 2019). We use this data product, as the GHS-BUILT has been used in a range of scientific studies of different disciplines (Ehrlich et al. 2021) and has been input to the multi-temporal population datasets GHS-POP and the rural-urban classification datasets GHS-SMOD (Florczyk et al. 2019). Moreover, GHS-BUILT R2018A makes use of early Landsat 4 MSS data and thus, extends farther back in time than related datasets such as the WSF-evolution dataset, which dates back to 1985 (Marconcini et al. 2020). For the two epochs 1975 and 2014, we extracted binary surfaces indicating built-up areas (1) and not built-up areas (0) (Figure 1a,b).

For the domain adaptation analysis, i.e., a test how predictive accuracy models perform on data of a different distribution than the one they were trained on (Section 2.2.6), we employed the GHS-BUILT-S2 dataset, which provides estimates of built-up probability, in the range of 0-100, within a 10x10m grid. GHS-BUILT-S2 has been created from Sentinel-2 data acquired in 2018, using convolutional neural networks (Corbane et al. 2021). We used the data for a subset of the city of Charlotte, North Carolina. For our accuracy assessment, these continuous data needed to be converted into binary, presence-absence surfaces. To do so, we calculated the average built-up probability of the 10m grid cells within 30x30m grid cells (aligned and consistent to the GHS-BUILT R2018A grid). We then applied a threshold of 50 to the built-up probabilities to generate binary built-up surface layers (Figure 3g), compatible to the GHS-BUILT R2018A data. However, since the data stem from a different sensor, resolution, processing strategy, and encoding, this ensures that this dataset represents a different data distribution than the GHS-BUILT R2018A (see also Uhl and Leyk 2022). This is done for two reasons: (a) the subsequent data processing requires binary, presence-absence surfaces, and (b) the original resolution of 10x10m is likely too fine-grained for direct calculation of landscape metrics. A target resolution of 30x30m generalizes the data such that meaningful landscape metrics can be derived (e.g., a contiguous patch of built-up surface should encompass the roads separating the actual buildings within that patch, and this may not be the case when using the original resolution of 10x10m).

*2.1.2. Gridded reference data*

The reference dataset has been created from the MTBF-33 vector building footprint data. MTBF-33 contains over 6 million building footprint vector geometries annotated with their construction year. For each county in the state of Massachusetts, we selected the MTBF-33 building footprints built-up by 1975, and 2014, respectively, and rasterized the vector data into the GHS-BUILT R2018A grid. To keep resampling uncertainty to a minimum, we first rasterized the vector polygons into a binary grid of 2x2m, and then down-sampled this grid to the target resolution of 30x30m, labelling all 30m grid cells as "built-up" if they contain at least one 2m building grid cell. A subset of these gridded surfaces is shown in Figure 1a,b). For the domain adaptation analysis, we carried out the same processes for the Mecklenburg County (i.e., the city of Charlotte, North Carolina) building footprints, but for the year 2016 only (Figure 3g).

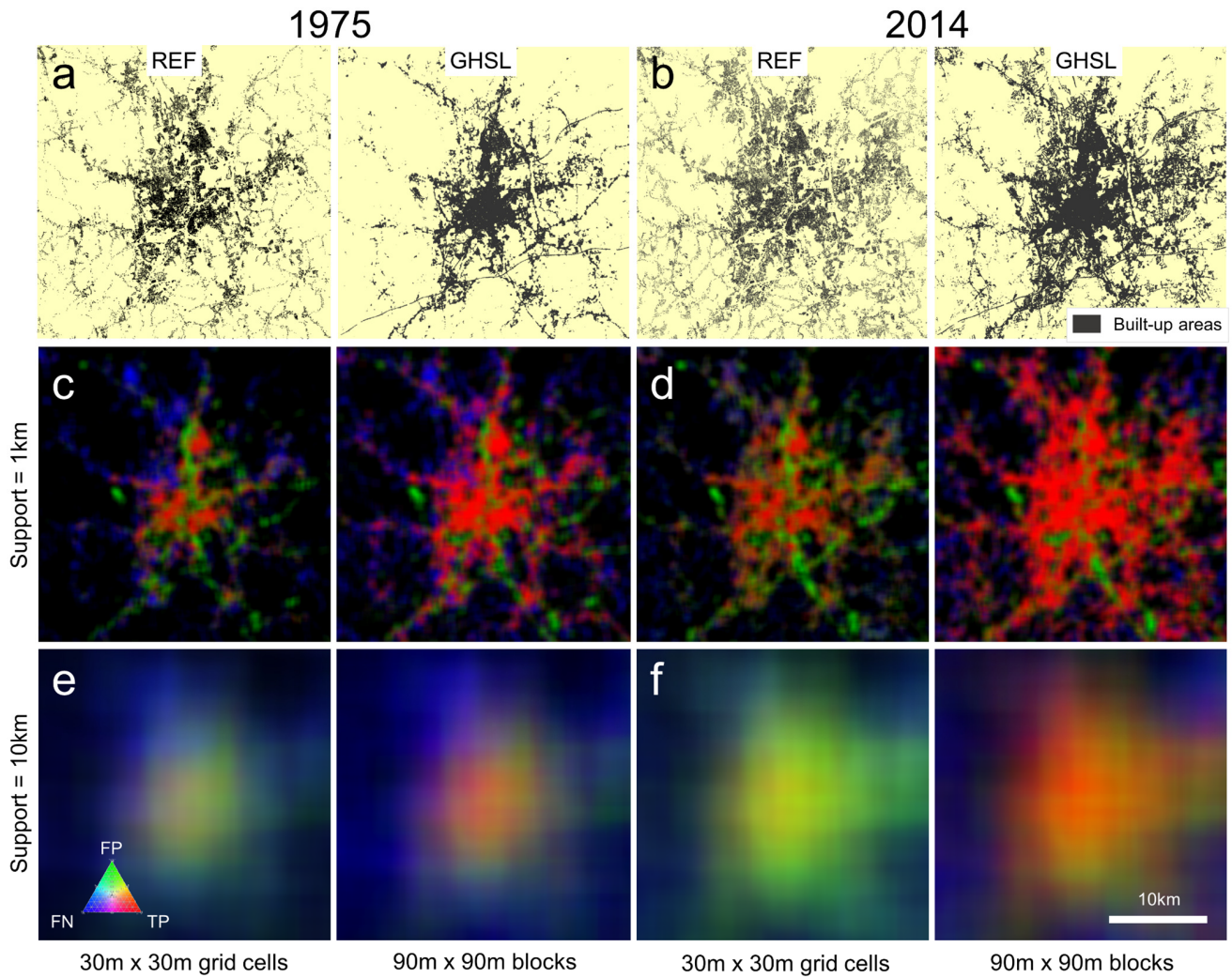

**Figure 1. Illustrating the input data, and derived focal confusion matrix composites for systematically varied parameters used in this study: Top row: Reference built-up surface layer derived from the Multi-temporal building footprint database MTBF-33, and test data from the GHS-BUILT R2018A (a) for the epoch 1975, and (b) for the epoch 2014. Middle row: Focal confusion matrix composites derived for the 1975 epoch at a support level of 1x1km for**

analytical units of 30x30m, and 90x90m, (c) in 1975, and (d) in 2014. Bottom row: Focal confusion matrix composites derived for the 1975 epoch at a support level of 10x10km for analytical units of 30x30m, and blocks of 90x90m, (e) in 1975, and (f) in 2014. Focal confusion matrix composites are RGB-encoded, i.e., the relative frequencies of the agreement categories are illustrated by the color tones. Specifically, true positives (TP) are represented by the red channel, false positives (FP) by the green channel, and false negatives (FN) by the blue channel. Thus, the colors provide a qualitative insight on the locally "dominating" agreement category and allows to visually detecting regions of high levels of agreement (red) or disagreement (blue for omission, green for commission errors). Data shown for the city of Worcester, Massachusetts, USA.

## 2.2. Methods

Herein, we used the pre-processed GHS-BUILT layer as test data and applied the MTBF-33 (Section 2.1) as reference data. Our method consists of the following steps: Spatially explicit map comparison (Section 2.2.1) and calculation of localized accuracy estimates (Section 2.2.2), the derivation of focal landscape metrics of built-up areas from both the reference and GHS-BUILT data (Section 2.2.3), the correlation analysis of localized accuracy and landscape metrics (Section 2.2.4) and predictive uncertainty modelling (Section 2.2.5), and finally, assessing the sensitivity of these results to the spatial support, to the epoch, to the analytical unit, and to the study area (Section 2.2.6). This workflow is shown in Figure 2.

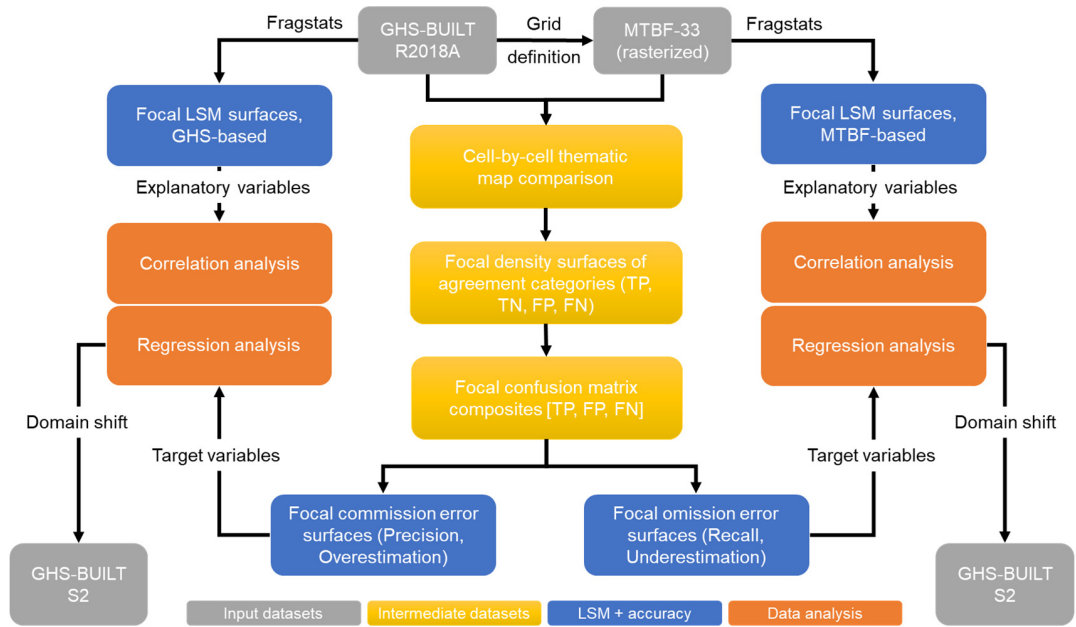

**Figure 2. Processing workflow for this study.**

### 2.2.1. Spatially explicit, exhaustive accuracy assessment

Based on the binary built-up presence/absence layers (Figure 1a,b) we applied a method for efficient, spatially explicit accuracy assessment of categorical, gridded data, as proposed in Uhl & Leyk (2022b). This method first performs cell-by-cell map comparison and generates three gridded surfaces, each one containing a 1-hot encoding of one of the three relevant agreement classes (i.e., true positives, false positives, false negatives). Subsequently, the densities of each agreement class within focal windows of varying size (herein called the "spatial support") are calculated. Finally, these agreement class density surfaces are stacked cell-wise to a three-band focal confusion matrix composite, representing the localized confusion matrix at each location (i.e., grid cell). Moreover, we needed to account for potential effects of positional uncertainty in our data, that may cause misalignment between GHS-BUILT and reference data, and could severely bias the thematic accuracy estimates obtained at the "native" resolution of 30x30m (e.g., Congalton 2007, Gu & Congalton 2020). To mitigate such effects, we down-sampled the binary GHS-BUILT and reference grids to blocks of 3x3 pixels (i.e., corresponding to a resolution of 90x90m) and repeated the steps described above, for the 90x90m grids, as well as for both epochs (i.e., 1975 and 2014). Finally, we expected our focal accuracy estimates to be sensitive to the spatial support (Uhl & Leyk 2022b), and thus, we used focal windows of varying size $s$ (1km, 2.5km, 5km, and 10km) to compute the agreement class density surfaces. Examples of the resulting confusion matrix composites for the different epochs, different levels of spatial support, and analytical units are shown in Figure 1c-f.

*2.2.2. Focal accuracy measures*

Based on the focal confusion matrix composites holding the densities of confusion matrix elements TP (true positives), FP (false positives), and FN (false negatives) (see Section 2.2.1, Figure 3a), we were able to efficiently calculate localized accuracy estimates at the grid-cell level. We calculated two thematic agreement metrics: Precision and Recall. Recall indicates the probability of a reference element being classified correctly, and is complementary to the omission error OE (error of exclusion), and the precision indicates the probability of a classified object being correct, and is complementary to the commission error CE (error of inclusion) (Story and Congalton 1986):

$$Recall = \frac{TP}{TP + FN} \tag{1}$$

and

$$Precision = \frac{TP}{TP + FP} \tag{2}$$

Besides these thematic agreement measures, we use the absolute error (AE), a quantity agreement measure, which is independent from the spatial correspondence of class labels within the focal windows. AE is obtained as:

$$AE = (TP + FP) - (TP + FN) = FP - FN \tag{3}$$

Where TP + FP represents the built-up quantity reported in GHS-BUILT, and TP + FN represents the built-up quantity according to the reference data (Figure 3b). These built-up quantities (i.e., the amount of 30x30m built-up grid cells per quadratic focal window of size $s$x$s$, given in meters) can be converted into a measure of built-up surface density (in %, herein called "built-up density"), as follows

$$BUDENS_{REF,s}[\%] = 100 \cdot 30^2 \cdot \frac{(TP + FN)}{s^2} \tag{4}$$

The GHSL-based built-up density is obtained as:

$$BUDENS_{GHSL,s}[\%] = 100 \cdot 30^2 \cdot \frac{(TP + FP)}{s^2} \tag{5}$$

Herein, we use these built-up density estimates to model the rural-urban continuum (Section 2.2.3). The aforementioned separation of thematic and quantity error has been proposed in a similar way by Pontius & Millones (2011), and allows to measure the agreement between test and reference data quantitatively, while ignoring the thematic agreement at the grid cell level. We use this strategy in regards to coarser-scale applications where the precise locations of built-up surfaces are irrelevant, for example when combining fine-scale built-up surface data with coarser population estimates. Examples of the resulting surfaces of thematic (i.e., precision and recall) and quantity agreement (i.e., AE) are shown in Figure 3c) for the different support levels and will be discussed in Section 3.3.

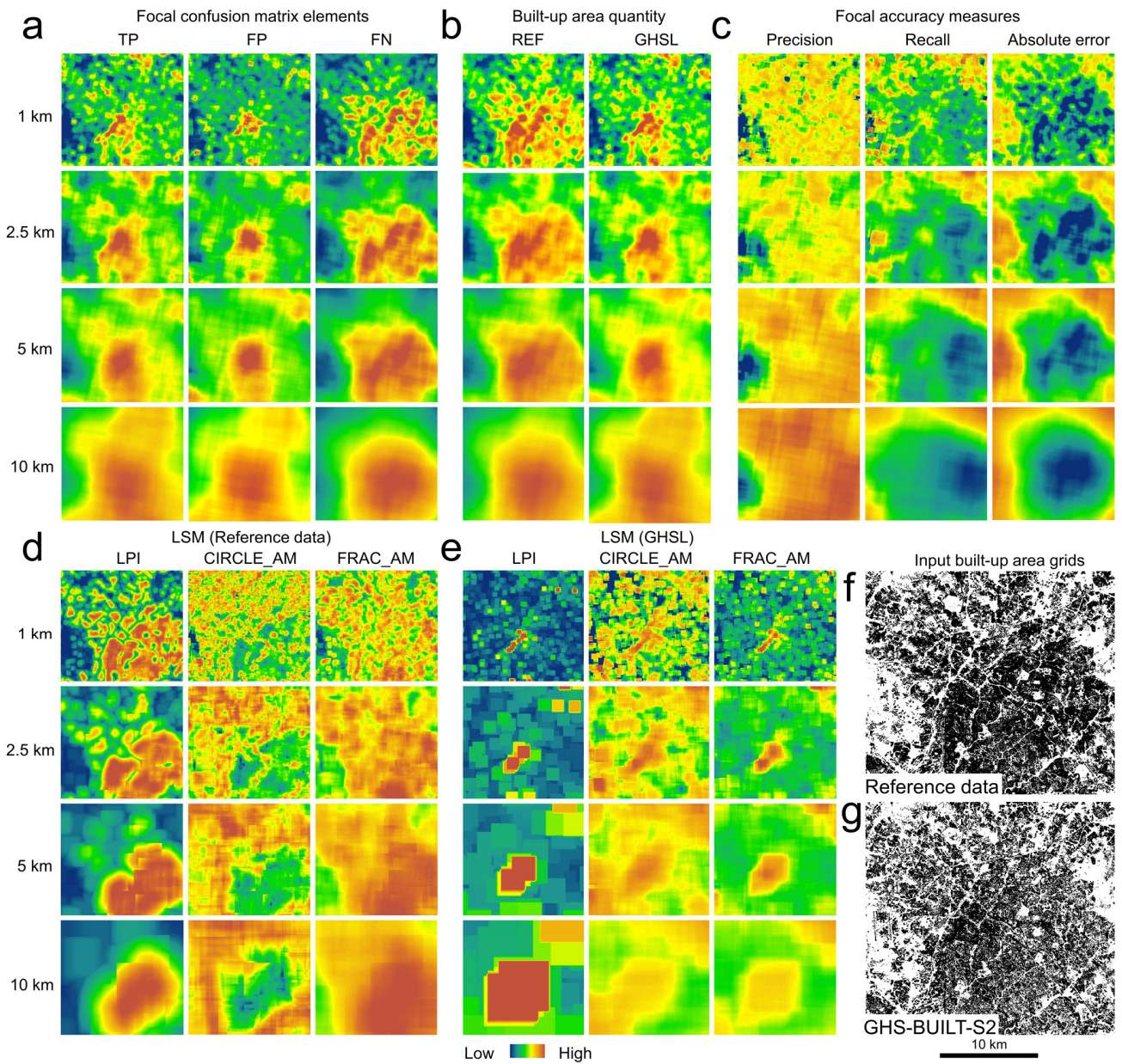

**Figure 3.** Illustrating the continuous surfaces used in this study and the effect of spatial support: (a) density surfaces of grid cells for each thematic agreement category (i.e., true positives – TP, false positives – FP, and false negatives – FN), (b) derived measures of built-up quantity (measured in grid cells) derived from the reference data and the test data, (c) surfaces of focal, thematic and quantity agreement measures, (d) selected focal landscape metrics (LSMs) derived from (d) the reference data, and (e) the GHS-BUILT-S2, (f) gridded built-up presence surface derived from the MTBF-33 reference database, and (g) corresponding test surface derived from aggregating and thresholding the GHS-BUILT-S2 built-up probabilities. All data shown for a subset of Charlotte, North Carolina. A rank transform was applied to the continuous surfaces before color-coding.

*2.2.3. Focal landscape metrics*

We used the software FRAGSTATS v4.2 (McGarigal et al. 2012) to calculate landscape metrics describing the shape and spatial structure of contiguous patches of built-up land within focal regions defined by the four support levels. To keep computational efforts manageable, we computed these metrics for a subset of N=200,000 locations within Massachusetts (see

Section 2.2.5). While previous work suggests that particularly the size of patches affects the classification accuracy (Smith et al. 2002 and 2003, Klotz et al. 2016, Mück et al. 2017), we also assume that certain shape and fragmentation characteristics may drive classification accuracy. Thus, we computed eight landscape level measures and eight patch-level measures, for both built-up areas from the GHSL and the reference data (Table 1). To characterize the distributions of all patch-level measures within the focal windows in Table 1, we calculated mean (MN), area-weighted mean (AM), median (MD), standard deviation (SD), coefficient of variation (CV), and range (RA), summing up to a total of 51 landscape metrics. As shown in Table 1, these commonly used metrics cover a wide range of the morphological, shape, and structure-related characteristics of built-up areas that are assumed to affect the classification accuracy of the GHSL in different ways. For our Charlotte study area (Figure 3f,g), we calculated exhaustive surfaces of focal landscape metrics, using grid and support levels consistent to the focal accuracy surfaces. These focal landscape metrics were derived from both the reference data (see Figure 3d for some examples) and the GHS built-up areas (Figure 3e) for the four levels of spatial support, as landscape metrics may be scale-sensitive (Lustig et al. 2015, Frazier 2022). The surfaces of all 51 landscape metrics, derived from the reference data, for the four support levels are shown in Appendix Figure A1.

**Table 1. Landscape metrics used in this study include 9 landscape-level measures, and 7 patch-level measures. For each patch-level measure, six summary statistics were computed. Source: McGarigal (2015).**

| Metric type | Metric name | Short name | Measured characteristic |
|---|---|---|---|
| Landscape / class level | Aggregation Index | AI | Disaggregation |
| | Landscape Division Index | DIVISION | Segregation |
| | Landscape Shape Index | LSI | Shape complexity |
| | Largest Patch Index | LPI | Dominance, connectivity |
| | Number or patches | NP | Segregation |
| | Percentage of Like Adjacencies | PLADJ | Contiguity |
| | Perimeter-area fractal dimension | PAFRAC | Shape complexity |
| | Edge density | ED | Compactness, shape complexity, segregation |
| | Cohesion index | COHESION | Connectivity |
| Patch level (MN, AM, MD, SD, CV, RA) | Contiguity Index | CONTIG | Contiguity |
| | Fractal Index | FRAC | Shape complexity |
| | Patch Area | AREA | Size |
| | Perimeter-Area Ratio | PARA | Shape complexity |
| | Radius of Gyration | GYRATE | Extension |
| | Related Circumscribing Circle | CIRCLE | Compactness |
| | Shape index | SHAPE | Shape complexity |

*2.2.4. Correlation analysis*

Using the consistent gridded layers of spatially corresponding focal accuracy estimates and the focal landscape metrics we can calculate the correlation between GHS-BUILT data accuracy and the morphological characteristics of the built-up areas. We calculated Pearson's correlation coefficient between the landscape metrics and focal accuracy estimates (Section 3.1), as well as between landscape metrics and built-up density, since we assume that built-up density may be a good proxy for GHS data accuracy, as previous work has shown (Leyk et al. 2018, Uhl & Leyk 2022b).

*2.2.5. Predictive uncertainty modeling*

We assessed the predictability of classification accuracy by means of landscape characteristics across different levels of spatial support. We drew two subsamples of N=100,000 from the initial sample through random selection, stratified by deciles of $BUDENS_{REF}$ (sample I) and $BUDENS_{GHSL}$ (sample II), respectively. This stratification ensured that the drawn samples are equally distributed across the rural-urban continuum and kept computational efforts feasible. For the locations in sample I, we computed focal landscape metrics based on the built-up patches in the reference data, and for the locations in sample II, we used the GHSL-derived built-up patches to compute landscape metrics (Section 2.2.3). As described above, we separately assessed thematic accuracy and quantity agreement. These components are further separated into omission and commission errors. While the reference data alone are independent from the test data (i.e., the GHSL), landscape metrics (LSMs) derived from the reference data ($LSM_{REF}$) neither contain any information on commission errors in the test data, nor do GHSL-based landscape metrics ($LSM_{GHS}$) allow for inferring on omission errors with respect to the reference data. Thus, we used precision and recall as response variables to be estimated based on the $LSM_{GHS}$ and, $LSM_{REF}$ respectively. Accordingly, we separated the absolute error (Equation 3) into overestimation (OE) and underestimation (UE) components as follows:

$$OE = \begin{cases} AE, & AE > 0 \\ 0, & AE \leq 0 \end{cases} \tag{4}$$

$$UE = \begin{cases} 0, & AE > 0 \\ abs(AE), & AE \leq 0 \end{cases} \tag{5}$$

Thus, we established four models: (a) Estimating thematic commission error, i.e. the precision of GHSL given the reference data, based on GHSL-derived landscape metrics:

$$Precision_{GHSL \leftarrow REF} = f(LSM_{GHSL}) \tag{6}$$

(b) Estimating quantity commission error, i.e. the OE of GHSL given the reference data, based on GHSL-derived landscape metrics:

$$OE_{GHSL \leftarrow REF} = f(LSM_{GHSL}) \tag{7}$$

(c) Estimating thematic omission error, i.e. the recall of GHSL given the reference data, based on reference-data derived landscape metrics:

$$Recall_{GHSL \leftarrow REF} = f(LSM_{REF}) \tag{8}$$

and (d) Estimating quantity omission error, i.e. the OE of GHSL given the reference data, based on reference-data derived landscape metrics:

$$UE_{GHSL \leftarrow REF} = f(LSM_{REF}) \tag{9}$$

All models were implemented as regression models using an AdaBoost regressor (Freund and Schapire 1997, Drucker 1997), which consists of an ensemble of shallow decision trees ("weak learners") and has shown promising performance in other applications in the geosciences (e.g., Li et al. 2016, Belgiu and Dragut 2016). We compared these model outcomes to a classical ordinary least squares (OLS) linear regression as a baseline model, in order to test whether complex machine learning models such as the AdaBoost regressor are necessary to solve the given regression problem, or if classical, and more interpretable statistical models such as OLS are sufficient. All models were tested using these two techniques and separately for the four levels of spatial support, in order to assess cross-scale effects, yielding a total of 32 regression models (Section 3.5). For the AdaBoost regression, we performed hyperparameter tuning separately for each response variable and support level. The outcomes of this analysis will illuminate two questions: (a) What can landscape metrics derived from the reference data tell us about omission errors in built-up land reported in the GHS-BUILT? (b) Can the GHS-BUILT itself be used to estimate its inherent uncertainty (i.e., commission errors)?

*2.2.6. Sensitivity analyses*

In our analytical setup, there are four components potentially affecting the performance of the predictive uncertainty models and the drawn conclusions. These components include:

1) The **spatial support** of localized accuracy estimates and landscape metrics. To address that, we carried out all regression models based on the four levels of spatial support and compared their results.
2) The **analytical unit** (i.e., the grid cell size). As mentioned before, positional uncertainty in our data may cause misalignment between the gridded GHS-BUILT and reference data, and could severely bias the thematic accuracy estimates obtained at the "native" resolution of 30x30m (Congalton 2007). Thus, we also computed the landscape metrics and localized accuracy estimates in coarser, 90x90m grids and carried out the regression analysis accordingly. This step is important because the effect of positional uncertainty on thematic accuracy estimates itself appears to depend on the landscape characteristics (Gu & Congalton 2020).
3) The **epoch** or acquisition date. As GHS-BUILT R2018A is a multi-temporal data product (1975-2014) using multispectral data from various generations of the Landsat sensors as input, the relationship between classification accuracy and underlying landscape metrics may vary over time, as the properties and capabilities of the underlying sensors (Landsat MSS, TM, ETM+, OLI) have changed over time. Thus, we also computed the landscape metrics and localized accuracy estimates based on the 1975 GHSL epoch, and on a 1975 snapshot of the MTBF-33 reference data.
4) The **study area and data product**. Landscape metrics may be very specific to the settlement patterns in Massachusetts. Moreover, the GHS-BUILT accuracy may be dependent of vegetation types, predominant roof material, and potentially ambiguous spectral responses between built-up and not built-up landscape features. Moreover, cloud cover frequency associated with a specific study area may also affect the GHS accuracy in that area. To account for that, we applied our regression models developed using the Massachusetts data to our Charlotte, North Carolina study area. For that study area, which is also covered by the MTBF-33 reference database, localized accuracy estimates and landscape metrics were derived from the fine-resolution GHS-BUILT-S2 product (see Section 2.1.1).

In the subsequent analyses, we considered the 2014 epoch and the analytical unit of 30x30m as our baseline scenario.

## 3. Results and discussion

In this section, we first describe the results of the correlation analysis between focal accuracy estimates and landscape metrics (Section 3.1), and the regression-based predictive uncertainty models (Section 3.2). We then discuss the sensitivity of correlation and regression analysis to the GHS epoch (1975 and 2014) and to the analytical unit underlying the spatially explicit accuracy assessment and landscape metrics computation (Section 3.3). Finally, we describe the results of the domain adaptation analysis, i.e., applying the regression models trained on GHS-BUILT R2018A in Massachusetts to GHS-BUILT-S2 data in North Carolina (Section 3.4).

### 3.1. Correlation analysis

As a first step, we systematically analyzed the correlation coefficients between each of the response variables, i.e., thematic commission error (precision), quantity commission error (overestimation), thematic omission error (recall), and quantity omission error (underestimation), and all 51 landscape metrics (i.e., 9 landscape-based measures, and 6 summary statistics for each of the 7 patch-based measures) used as explanatory variables. Figure 4a shows the correlation coefficients for these metrics (for patch-based measures, only the statistic with maximum average correlation across all 16 models is shown, see Figure A2 for the full matrix, and Figure A3 for a more detailed visualization of the relationships between landscape metrics and selected accuracy measures). The landscape metrics are sorted by their average correlation to all accuracy metrics at all support levels. We observe the following: In average, the Landscape shape index (LSI) exhibits the highest levels of correlation to the accuracy measures under test. Among the tested landscape metrics, highest levels of correlation are found for the reference-based landscape metrics (LSM$_{REF}$) and recall, and lowest for UE. In many cases, correlation increases with increasing spatial support. As shown in Figure 4a, contiguity, disaggregation, and connectivity measures of built-up land (AI, PLADJ, COHESION, and CONTIG) are highly correlated with recall, whereas quantity omission errors (i.e., UE) are highly correlated with measures of shape complexity (ED, LSI) but also with scatteredness (i.e., NP, also ED). Among the GHSL-based landscape metrics (LSM$_{GHS}$) that strongly correlate with commission error measures (i.e., precision and OE) are the Shape index (SHAPE), Fractal index (FRAC) and the GYRATE metric, measuring shape complexity and extension, yielding higher values e.g. for irregular road features, where commission errors typically occur. Thus, commission errors appear to be associated with the shape of the GHSL built-up areas, whereas omission errors are related to the contiguity and segregation of reference built-up areas.
Moreover, we visualized the 51 landscape metrics and the accuracy components in a bi-dimensional space defined by their cross-support correlation trajectory with respect to commission error measures, and omission error measures (Figure 4b). This way of visualizing the results shows highest correlations between landscape metrics and thematic omission error (i.e., recall) and most of these metrics also are highly correlated to thematic commission errors (i.e., precision). Conversely, some of the quantity agreement measures exhibit higher levels of correlation to OE, but low correlation to UE, indicating that structural properties of built-up areas determining the level of quantity overestimation are different from those that trigger underestimation.

### 3.2. Predictive uncertainty modeling

Furthermore, we visualized the performance of the predictive AdaBoost models generated for each of the four scenarios (Section 2.2.6), separately for models predicting thematic accuracy (Figure 4c) and quantity agreement (Figure 4d). These visualizations show that for the optimal hyperparameters, all four models yield $R^2$ values of >0.9, indicating that all four accuracy components can be predicted reliably based on the landscape metrics. Visualizing RMSE versus $R^2$ for each model reveals further that there are increasing levels of $R^2$ as spatial support increases. The opposite trends between Figure 4c and d along the x-axis are due to the absolute nature of the OE and UE quantity error components, naturally increasing with increasing spatial support. As can be seen, the best-fitting models are achieved for predicting recall from $LSM_{REF}$ and for predicting OE from $LSM_{GHS}$.

These observations imply that landscape metrics derived from the GHSL can reliably be employed as predictors of commission errors in the absence of reference data, and the omission error component of the accuracy of built-up land data is highly affected by the level of spatial segregation and contiguity of built-up areas, confirming prior studies (e.g., Smith et al. 2002 and 2003, Klotz et al. 2016, Mück et al. 2017). Overall, recall appears to be highest correlated to LSMs, and models exhibit the highest predictive power. It is also worth noting that while the machine-learning models (AdaBoost) consistently outperform the OLS models in most cases, OLS comes closest to the AdaBoost model performance for predicting recall for large spatial supports (Table 2).

**Table 2. Regression analysis results for the predictive modelling of GHS accuracy based on landscape metrics, using AdaBoost regression and Ordinary Least Squares.**

| Landscape metric source data | Response variable | Spatial support [m] | Max. Depth | Num. Estimators | AdaBoost Regressor R2 | RMSE | OLS R2 | RMSE |
|---|---|---|---|---|---|---|---|---|
| GHSL | OE | 1000 | 10 | 500 | 0.342 | 57.425 | 0.377 | 55.198 |
| | | 2500 | 25 | 250 | 0.575 | 296.407 | 0.545 | 306.091 |
| | | 5000 | 25 | 500 | 0.800 | 722.664 | 0.653 | 940.203 |
| | | 10000 | 25 | 250 | 0.949 | 1151.146 | 0.751 | 2531.439 |
| GHSL | Precision (User's Acc.) | 1000 | 10 | 500 | 0.119 | 0.141 | 0.129 | 0.142 |
| | | 2500 | 25 | 250 | 0.412 | 0.116 | 0.342 | 0.123 |
| | | 5000 | 25 | 500 | 0.696 | 0.079 | 0.447 | 0.108 |
| | | 10000 | 25 | 500 | 0.913 | 0.039 | 0.539 | 0.090 |
| Reference data | UE | 1000 | 10 | 500 | 0.268 | 37.740 | 0.264 | 37.796 |
| | | 2500 | 25 | 500 | 0.550 | 107.468 | 0.442 | 120.762 |
| | | 5000 | 25 | 250 | 0.791 | 209.211 | 0.525 | 315.497 |
| | | 10000 | 25 | 500 | 0.909 | 358.610 | 0.511 | 838.903 |
| Reference data | Recall (Prod. Acc.) | 1000 | 10 | 250 | 0.573 | 0.174 | 0.561 | 0.177 |
| | | 2500 | 25 | 500 | 0.819 | 0.119 | 0.781 | 0.133 |
| | | 5000 | 25 | 500 | 0.928 | 0.073 | 0.857 | 0.104 |
| | | 10000 | 25 | 500 | 0.985 | 0.032 | 0.903 | 0.080 |

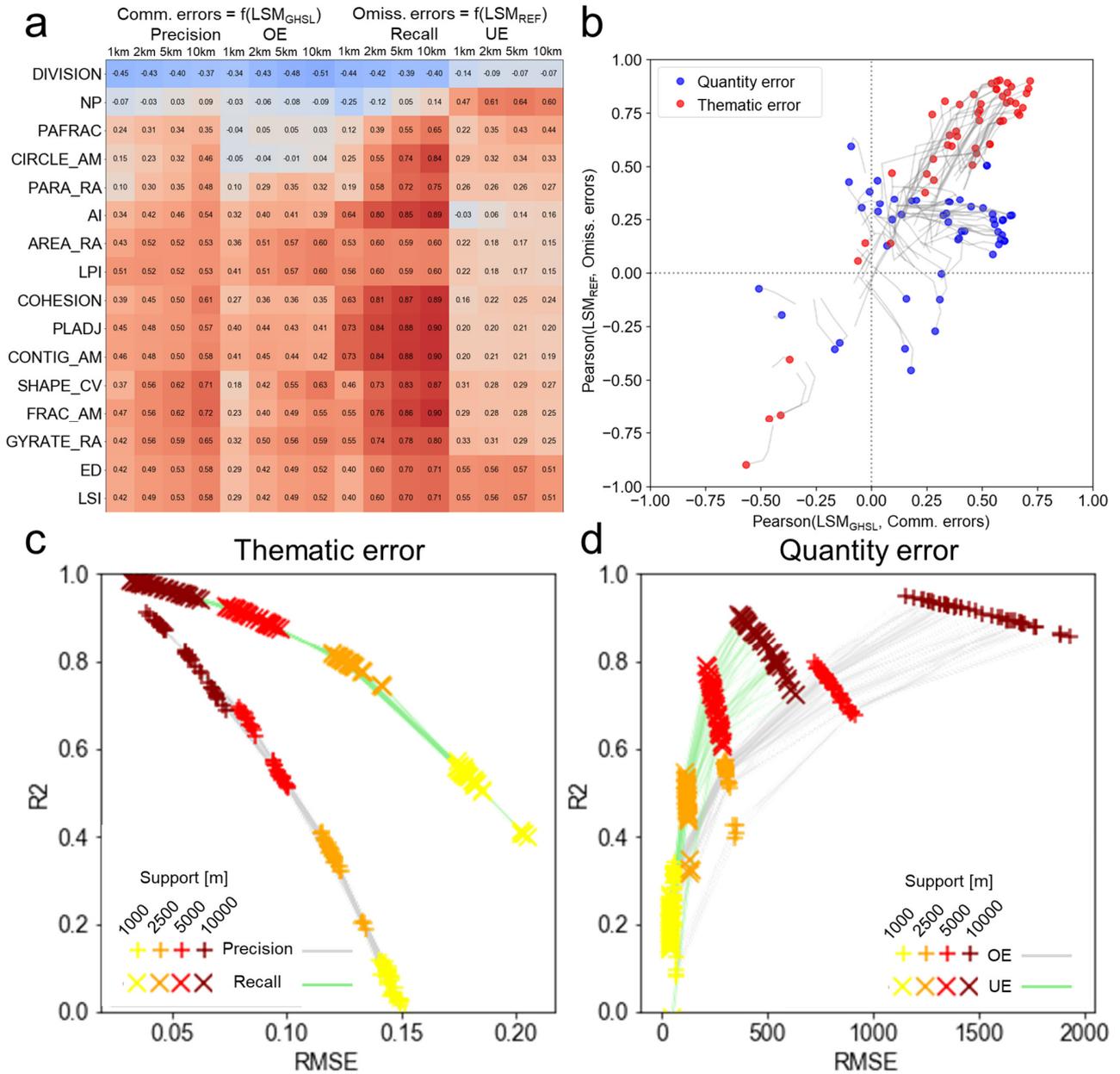

**Figure 4. Predictive analysis of localized accuracy using focal landscape metrics.** (a) Pearson's correlation coefficients for the 16 most correlated landscape metrics, for each response variable and each level of spatial support. For patch-based metrics, only the summary statistic is shown that yields the highest overall correlation, see Appendix Figure A2 for the full matrix; LSMs are sorted from top to bottom ascendingly by their average correlation across rows; (b) Correlation coefficients of the 51 landscape metrics in a bi-dimensional space of correlation to measures that characterize commission error (x-axis) and omission error (y-axis), color-coded by accuracy type (i.e., thematic or quantity agreement); correlation coefficients are shown for spatial support of 10km, grey lines illustrate the cross-support trajectory for each LSM across the four levels of spatial support; (c) shows the AdaBoost model performance in bi-dimensional spaces of RMSE and $R^2$ for the two models predicting thematic errors, with each data point representing a different hyperparameter setting; (d) respective visualization for the two models predicting quantity errors. Lines connect the R2-RMSE pairs for each hyperparameter combination across the levels of spatial support.

## 3.3. Sensitivity to epoch and analytical unit

Revisiting the focal confusion matrix composites shown for the epochs 1975 and 2014, and for the analytical units of 30x30m and 90x90m (Figue 1c,d), we observe interesting differences in the relative proportions of TP, FP, and FN instances (i.e. grid cells). For example, the RGB-encoding of these relative proportions yields green-yellow colors in the center of the map (i.e., the city of Worcester, Massachusetts) for the 30m scenario, and these areas turn red in the 90m scenario, indicating higher proportions of grid cells switching from false positive to true positive when using a coarser analytical unit. This effect could be due to actual misalignments, which are mitigated by the 90m aggregation, or could be caused by actual false positives (e.g., roads classified as built-up areas) nearby true positive grid cells. Moreover, in Figure 1c we observe a blue fringe around the city of Worcester in both 1975 scenarios, indicating higher levels of omissions in the GHS-BUILT epoch 1975 in peri-urban areas. These blue color tones are less pronounced in the 2014 scenarios, indicating a decrease of false negatives relative to the other categories (TP, FP).

These observations imply that classification accuracy varies considerably across GHSL epochs, and that the chosen analytical unit likely affects the magnitude of the resulting accuracy measures. How do these sensitivities affect the relationship between accuracy and landscape metrics, as measured by their correlation coefficients (Figure 4a, Appendix Figure A2)? To shed light on this question, we visualized the correlation coefficients for all landscape and accuracy metrics for the four scenarios (i.e., using epochs 1975 and 2014, respectively, at an analytical unit of 30m, and 90m, respectively, see Appendix Figure A4). While the overall trends seem persistent across these four scenarios, is the ranking of correlation coefficients, and thus the level of association between landscape and accuracy metrics consistent across scenarios? We transformed the correlation coefficients for each scenario in percentile-based ranks and visualized them in Q-Q plots (Figure 5).

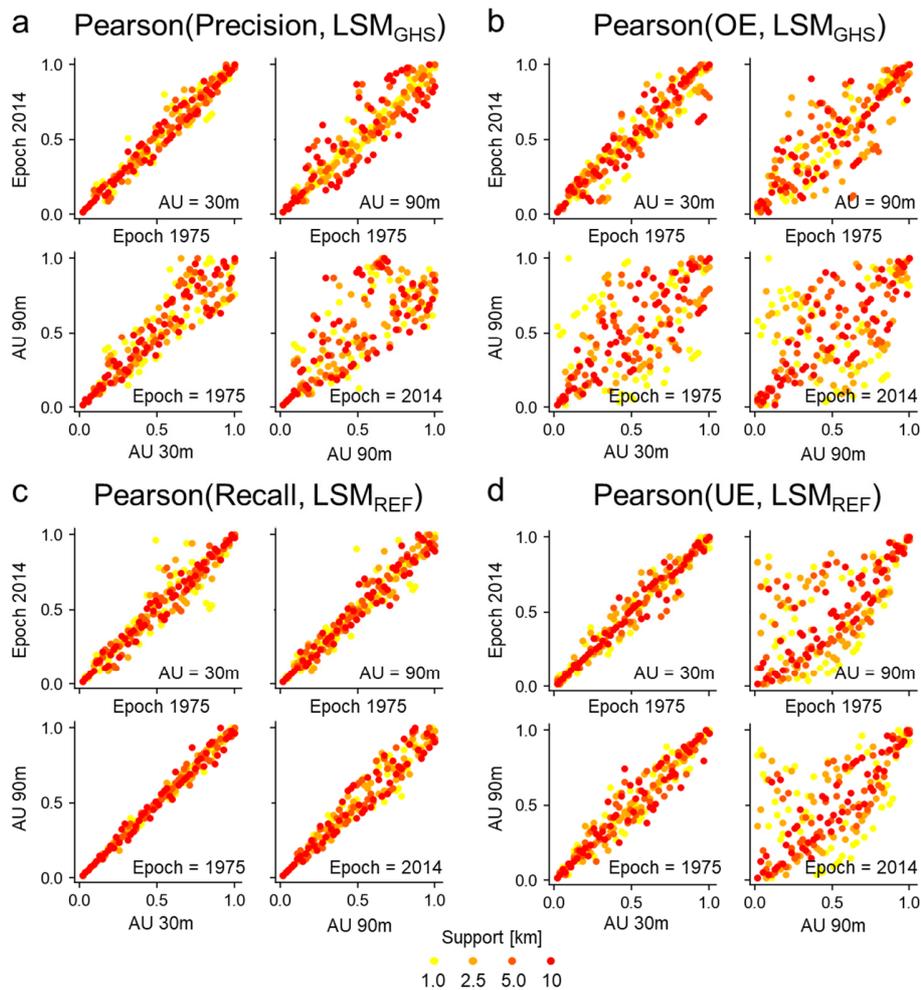

**Figure 5. QQ-plots of correlation coefficients between accuracy metrics and LSMs, for the four scenarios i.e., different GHSL epochs (i.e., 1975 and 2014), and different analytical units (AU; i.e., 30m, 90m) used for the accuracy assessments.**

The more spread the distributions in Figure 5 show, the more does either the epoch or the analytical unit (AU) affect the ranking of correlation coefficients. As can be seen in Figure 5c, the correlation coefficients between Recall and reference-data based landscape metrics experiences the least spread, with the points located nearby the main diagonal, indicating that the order of how strong the associations between specific landscape metrics and the Recall are, is largely independent from the GHSL epoch and from the chosen analytical unit. Conversely, the order of correlation coefficients between overestimation and GHSL-based landscape metrics is most affected by the epoch and analytical unit of the underlying data (Figure 5b).

The observed robustness of the correlation coefficients between individual landscape metrics and the recall in the GHS-BUILT built-up areas across epochs and analytical units, is also reflected in the regression analyses carried out for the four scenarios (Table 3). The $R^2$ values of all accuracy prediction models are relatively stable across the four scenarios. However, the coefficient of variation across the $R^2$ values of the OLS regression models that predict recall using $LSM_{REF}$ are considerably lower than for the other target variables. Importantly, the previously observed trend of increasing model fit with increasing spatial support (i.e., from 1km towards 10km) also persists when using the 1975 epoch or accuracy estimates obtained at 90m analytical unit. When looking at the average $R^2$ across the models for each of the four scenarios (bottom row of Table 3), we observe, on average, lowest model fits for the 1975 GHSL epoch and using an analytical unit of 90m. This drop in model fit is most pronounced when using GHS-based landscape metrics, indicating that the prediction of commission errors based on the GHS-BUILT alone is more difficult in 1975 than for the 2014 epoch.

**Table 3. Regression results across the four spatial support levels for GHSL epochs 1975 and 2014, and for analytical units of 30m and 90m.**

| LSM source | Spatial support [m] | Accuracy measure | RMSE per analytical unit and epoch | | | | $R^2$ per analytical unit and epoch | | | | $R^2$ Coefficient of variation |
|---|---|---|---|---|---|---|---|---|---|---|---|
| | | | 30m, 2014 | 30m, 1975 | 90m, 2014 | 90m, 1975 | 30m, 2014 | 30m, 1975 | 90m, 2014 | 90m, 1975 | |
| GHS | 1000 | OE | 90.999 | 93.864 | 9.241 | 11.622 | 0.298 | 0.245 | 0.191 | 0.114 | 0.320 |
| GHS | 2500 | OE | 372.065 | 390.618 | 37.702 | 49.065 | 0.421 | 0.362 | 0.247 | 0.203 | 0.284 |
| GHS | 5000 | OE | 1174.696 | 1214.789 | 122.756 | 139.083 | 0.464 | 0.427 | 0.201 | 0.279 | 0.313 |
| GHS | 10000 | OE | 833.944 | 2117.633 | 190.991 | 264.154 | 0.972 | 0.821 | 0.802 | 0.699 | 0.119 |
| Reference | 1000 | UE | 37.796 | 41.537 | 9.402 | 9.626 | 0.264 | 0.188 | 0.271 | 0.240 | 0.136 |
| Reference | 2500 | UE | 120.762 | 138.817 | 30.432 | 25.837 | 0.442 | 0.277 | 0.496 | 0.471 | 0.203 |
| Reference | 5000 | UE | 315.497 | 362.587 | 83.073 | 67.511 | 0.525 | 0.372 | 0.589 | 0.533 | 0.159 |
| Reference | 10000 | UE | 838.903 | 895.912 | 227.732 | 179.768 | 0.511 | 0.443 | 0.670 | 0.570 | 0.152 |
| GHS | 1000 | Precision | 0.128 | 0.122 | 0.163 | 0.196 | 0.363 | 0.383 | 0.298 | 0.258 | 0.153 |
| GHS | 2500 | Precision | 0.119 | 0.115 | 0.159 | 0.202 | 0.515 | 0.525 | 0.386 | 0.303 | 0.214 |
| GHS | 5000 | Precision | 0.099 | 0.099 | 0.134 | 0.189 | 0.591 | 0.580 | 0.461 | 0.282 | 0.259 |
| GHS | 10000 | Precision | 0.063 | 0.069 | 0.088 | 0.129 | 0.776 | 0.735 | 0.659 | 0.464 | 0.182 |
| Reference | 1000 | Recall | 0.177 | 0.184 | 0.203 | 0.237 | 0.561 | 0.526 | 0.480 | 0.391 | 0.130 |
| Reference | 2500 | Recall | 0.133 | 0.146 | 0.152 | 0.185 | 0.781 | 0.734 | 0.745 | 0.655 | 0.063 |
| Reference | 5000 | Recall | 0.104 | 0.116 | 0.119 | 0.147 | 0.857 | 0.822 | 0.841 | 0.770 | 0.040 |
| Reference | 10000 | Recall | 0.080 | 0.089 | 0.087 | 0.107 | 0.903 | 0.880 | 0.900 | 0.860 | 0.020 |
| | Average | | | | | | 0.578 | 0.520 | 0.515 | 0.443 | |

## 3.4. Domain adaptation analysis

Finally, we investigate how our OLS and AdaBoost regression models perfom when deployed on data from a different distibution. This is called domain shift, and models that yield good results when performing a domain shift, are capable of domain adaptation (You et al. 2019). To do so, we apply the models trained on the GHS-BUILT R2018A sample collected in Massachusetts, to a region in Charlotte, North Carolina, where focal accuracy and GHS-based landscape metrics were obtained from the GHS-BUILT-S2 product (see Sections 2.2.1 and 2.2.6). We visually compare three accuracy surfaces: (a) the calculated accuracy surfaces based on map comparison between GHS-BUILT-S2 and the MTBF-33 reference data, (b) the accuracy surfaces as predicted by the regression model trained on Massachusetts data (i.e., domain shift), and (c) the accuracy surfaces as predicted by a regression model trained on 80% of the data based on GHS-BUILT-S2 in the Charlotte study area (i.e., no domain shift). These surfaces are shown in Figure 6, for all target variables, support levels, and for the two regression techniques. As can be seen, the predicted accuracy surfaces using domain shift differ, in many cases, considerably from the calculated surfaces. In some cases (e.g., OLS-based recall prediction at a spatial support of 1km and 5km) the resulting surfaces are even inverted, indicating that the underlying relationships between specific landscape metrics and data accuracy may be inverted between the Landsat-based GHS-BUILT R2018A and the GHS-BUILT-S2 product, in the analyzed study area. Importantly, the OLS-based regression models perform the domain shift at a spatial support of 10km for most target variables, in particular for the models predicting precision and recall measures ($R^2$ of 0.42, and 0.46, respectively, highlighted in grey in Figure 6). Poor performance for the overestimation models is due to predominant built-up quantity underestimation in our Charlotte study area, and the resulting sparsity of focal regions where quantity overestimation occurs, impede the successful prediction.

Moreover, we observe that at a spatial support of 10km, OLS-based models appear to outperform AdaBoost regression models (e.g., three out of four OLS models show – visually - acceptable domain shift results at a support level of 10km, whereas this is not the case for any of the target variables using AdaBoost regression). This is in contrast to the better model fits of AdaBoost compared to OLS in Table 2, and indicated that the AdaBoost models may be overfitted to the Massachusetts study area, whereas the OLS-based models, despite exhibiting lower levels of model fit in the Massachusetts study area, appear to be more generalizable to other study areas, when the spatial support is large enough.

These results indicate that the morphological landscape characteristics that drive the presence or absence of both thematic commission and omission errors are largely identical for the GHS-BUILT R2018A and the GHS-BUILT-S2 product, indicating the potential for reliable accuracy prediction across data products, given that sufficient spatial context is provided.

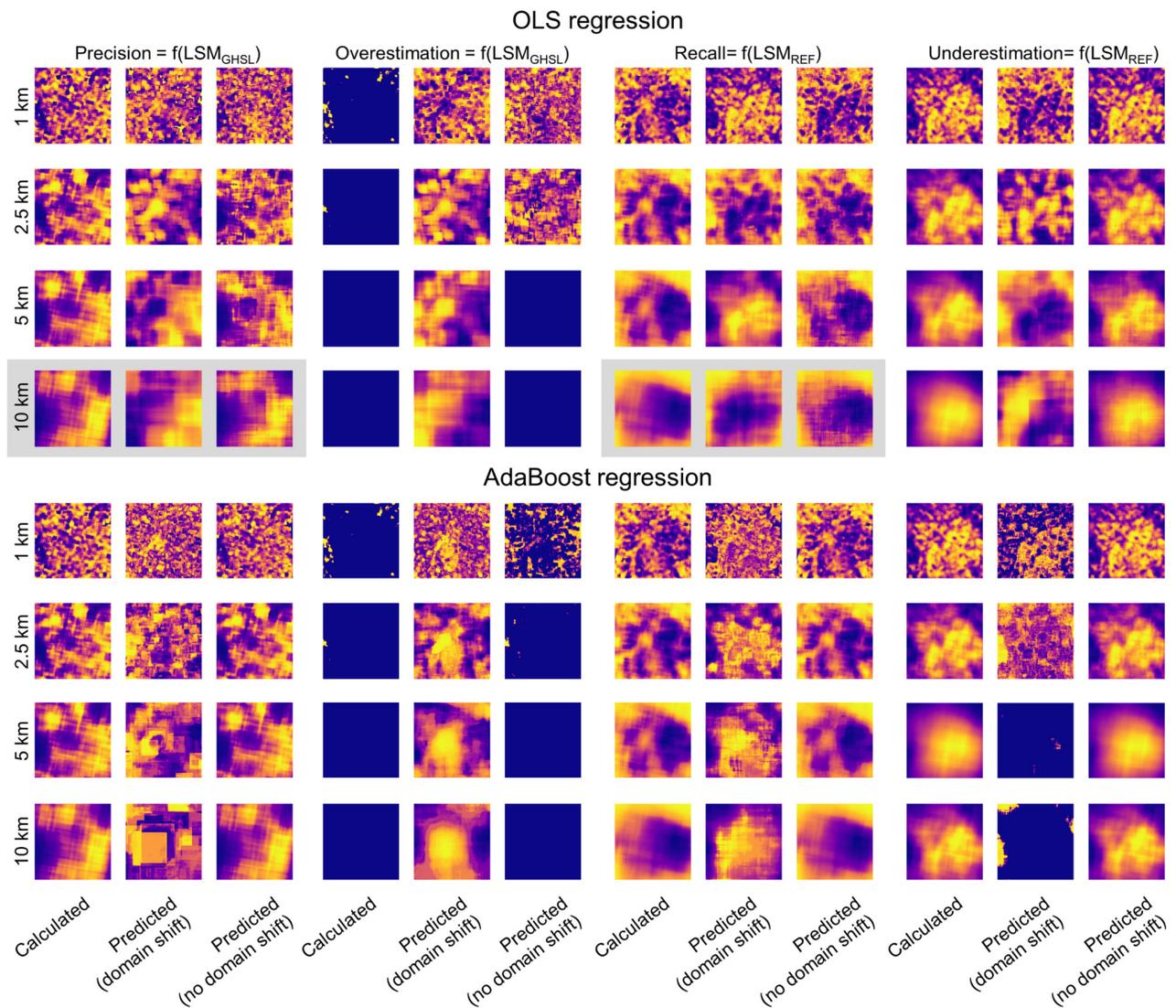

**Figure 6. Results of the domain adapatation tests for OLS and AdaBoost regression. Best domain adaptation results are achieved for predicting focal precision and recall using an OLS regression model at 10km spatial support (highlighted in grey). Values are rank-transformed; high values shown in yellow.**

While the presented analysis focused on the state of Massachusetts, we have calculated focal landscape metric surfaces based on the MTBF-33 reference data for all 33 counties covered by MTBF-33. In previous work, we have shown that there are strong associations between GHSL data accuracy and the density of built-up surface within a given spatial unit (Uhl & Leyk 2022b). Thus, we calculated the correlation coefficients between each landscape metric and built-up density for each of the 33 counties, and for three levels of spatial support (i.e., 1km, 2.5km, and 5km, see Appendix Figure A5). As can be seen, across the three levels of support, most landscape metrics exhibit high positive of negative correlation with built-up density, and these correlations are very consistent across the 33 counties, out of which 19 are located outside of the state of Massachusetts. Such a uniform picture across different geographic regions gives further confidence about the high levels of generalizability of our models.

## 4. Conclusions

In this article, we conducted a detailed assessment of the relationships between morphological characteristics of built-up surfaces (measured by means of landscape metrics), and the data accuracy of built-up areas reported in the gridded, multi-temporal GHS-BUILT R2018A dataset. We identified strong associations between accuracy measures and morphological characteristics of built-up areas, and high predictive power in the accuracy models, separately for omission and commission errors. The identified associations enable the effective creation of uncertainty models by means of proxy variables (such as landscape metrics), enabling localized, uncertainty-informed settlement modelling. These findings are useful to determine areas where omission errors are expected to be high, and could be incorporated into classifier training procedures, in order to improve future settlement layers. Moreover, the presented accuracy prediction models can be applied to existing built-up land data, to identify regions where commission errors are expected to be high, in the absence of reference data, and could inform the sampling design of future accuracy assessments.

While the tree-based AdaBoost regressor outperformed the OLS models in the "baseline scenario" (i.e., for the epoch 2014, using the full analytical resolution of 30x30m grid cells), our domain adaptation analysis revealed that these AdaBoost models likely overfitted to the Massachusetts study area, as they performed poorly in the "unseen" Charlotte study area. This important insight highlights the importance of domain shift / domain adaptation analyses when evaluating machine learning models, and revealed that our OLS-based models are highly generalizable for accuracy prediction in a different study area and using a different underlying data product. Furthermore, our analyses revealed that in particular our findings on the prediction of thematic omission error (i.e., the recall) are highly generalizable across different epochs of the GHS-BUILT, and are largely unaffected by the choice of the underlying analytical unit.

Notably, both correlations and model fits increased with the level of spatial support, indicating that the choice of an appropriate level of spatial support is crucial when creating and analyzing localized accuracy estimates and local landscape metrics. These observations underline the importance of scale-related considerations in geospatial analyses. However, which level of spatial support is appropriate for a specific purpose needs to be decided for each individual case, taking into account the tradeoff between model robustness (which increases with increasing support level in this study) on the one hand, and loss of spatial granularity on the other hand.

At this point it is important to mention that despite the domain adaptation analysis presented in Section 3.4, further work using a larger set of study areas is required to formalize general guidelines on the effects of landscape characteristics and GHS-BUILT data accuracy. In future work, we will also focus on the application of the described framework to different built-up surface / settlement data products and we will analyze in detail the sensitivity of landscape metrics to spatial support, taking into account potential bias introduced by the scale sensitivity of the landscape metrics themselves (see Lustig et al. 2015). While the relationships between landscape characteristics and data accuracy have been studied in the case of land cover data in general (Smith et al. 2002 and 2003), and, in the case of built-up land data (Klotz et al. 2016, Mück et al. 2017), this work demonstrated at unprecedented depth, that the accuracy of remote-sensing derived built-up land data products such as the GHS-BUILT is affected by the morphology of the built-up area patterns. Concluding, this work contributes to a better understanding of the spatial structure and variation of the uncertainty inherent in data products such as the GHS-BUILT R2018A, and ultimately, to a more informed and reflected use of such data products.

**Acknowledgements:** Funding for this work was partially provided through the National Science Foundation (awards 1924670 and 2121976 to CU Boulder). This research benefited from support provided to the University of Colorado Population Center (CUPC, Project 2P2CHD066613-06) from the Eunice Kennedy Shriver Institute of Child Health Human and Human Development. The content is solely the responsibility of the authors and does not necessarily represent the official views of the National Institutes of Health or CUPC.

**Data and code availability:** Focal landscape metrics and focal accuracy estimates computed for the epochs 1975 and 2014, for analytical units of 30x30m and 90x90m, as well as for the four levels of spatial support are available at https://doi.org/10.6084/m9.figshare.19785877. Python code for spatially explicit accuracy assessments of binary, gridded datasets is available at https://github.com/johannesuhl/local_accuracy.

## 6. Appendix

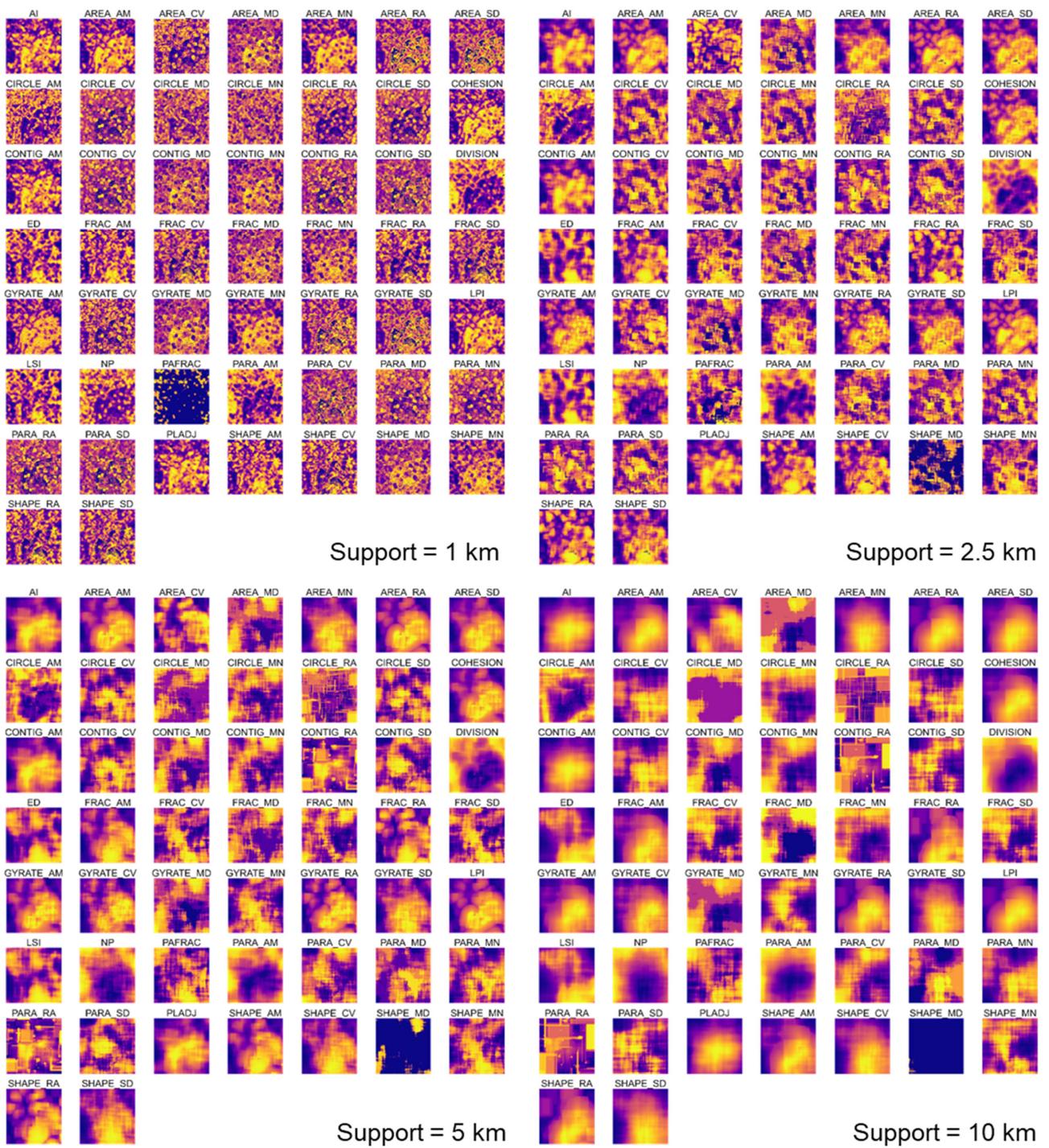

**Figure A1.** Exhaustive focal landscape metric surfaces at various levels of spatial support, shown for the city of Charlotte, North Carolina. Values are rank-transformed; high values shown in yellow.

| | Comm. errors = f(LSM_GHS) | | | | | | | | Omiss. errors = f(LSM_REF) | | | | | | | |
|---|---|---|---|---|---|---|---|---|---|---|---|---|---|---|---|---|
| | Precision | | | | OE | | | | Recall | | | | UE | | | |
| | 1km | 2km | 5km | 10km | 1km | 2km | 5km | 10km | 1km | 2km | 5km | 10km | 1km | 2km | 5km | 10km |
| PARA_AM | -0.44 | -0.47 | -0.49 | -0.57 | -0.40 | -0.44 | -0.43 | -0.41 | -0.72 | -0.84 | -0.88 | -0.90 | -0.20 | -0.20 | -0.21 | -0.20 |
| PARA_MN | -0.34 | -0.30 | -0.32 | -0.46 | -0.26 | -0.24 | -0.17 | -0.17 | -0.47 | -0.53 | -0.62 | -0.68 | -0.12 | -0.24 | -0.34 | -0.36 |
| DIVISION | -0.45 | -0.43 | -0.40 | -0.37 | -0.34 | -0.43 | -0.48 | -0.51 | -0.44 | -0.42 | -0.39 | -0.40 | -0.14 | -0.09 | -0.07 | -0.07 |
| PARA_MD | -0.30 | -0.26 | -0.29 | -0.41 | -0.23 | -0.17 | -0.11 | -0.14 | -0.42 | -0.49 | -0.60 | -0.66 | -0.12 | -0.21 | -0.29 | -0.33 |
| CIRCLE_CV | -0.08 | -0.01 | -0.02 | -0.06 | -0.05 | 0.03 | 0.13 | 0.15 | -0.05 | 0.09 | 0.11 | 0.06 | -0.02 | -0.12 | -0.26 | -0.35 |
| CONTIG_CV | -0.04 | 0.04 | 0.02 | -0.03 | -0.02 | 0.09 | 0.19 | 0.18 | 0.02 | 0.18 | 0.19 | 0.14 | -0.03 | -0.19 | -0.35 | -0.45 |
| CIRCLE_SD | -0.04 | 0.07 | 0.10 | 0.09 | -0.07 | 0.00 | 0.12 | 0.16 | -0.02 | 0.24 | 0.39 | 0.47 | 0.11 | 0.03 | -0.07 | -0.12 |
| NP | -0.07 | -0.03 | 0.03 | 0.09 | -0.03 | -0.06 | -0.08 | -0.09 | -0.25 | -0.12 | 0.05 | 0.14 | 0.47 | 0.61 | 0.64 | 0.60 |
| CIRCLE_RA | -0.05 | 0.08 | 0.16 | 0.28 | -0.07 | -0.05 | -0.01 | -0.01 | -0.12 | 0.13 | 0.35 | 0.44 | 0.28 | 0.35 | 0.40 | 0.38 |
| CIRCLE_MD | 0.10 | 0.11 | 0.12 | 0.25 | 0.00 | -0.01 | -0.05 | -0.04 | 0.17 | 0.27 | 0.38 | 0.47 | 0.15 | 0.21 | 0.28 | 0.30 |
| CIRCLE_MN | 0.11 | 0.10 | 0.14 | 0.28 | 0.01 | -0.04 | -0.11 | -0.10 | 0.15 | 0.23 | 0.40 | 0.54 | 0.18 | 0.30 | 0.40 | 0.43 |
| SHAPE_MD | 0.30 | 0.16 | 0.15 | 0.24 | 0.15 | 0.11 | 0.05 | 0.07 | 0.31 | 0.30 | 0.41 | 0.38 | 0.06 | 0.13 | 0.19 | 0.13 |
| AREA_MD | 0.30 | 0.10 | 0.05 | 0.34 | 0.19 | 0.08 | 0.01 | 0.09 | 0.24 | 0.06 | 0.51 | 0.60 | 0.03 | 0.19 | 0.27 | 0.25 |
| PARA_SD | 0.13 | 0.28 | 0.27 | 0.27 | 0.12 | 0.28 | 0.31 | 0.29 | 0.29 | 0.62 | 0.72 | 0.74 | 0.09 | -0.04 | -0.18 | -0.27 |
| FRAC_MD | 0.29 | 0.21 | 0.22 | 0.36 | 0.15 | 0.09 | 0.01 | 0.03 | 0.36 | 0.41 | 0.53 | 0.60 | 0.12 | 0.22 | 0.28 | 0.29 |
| GYRATE_MD | 0.32 | 0.15 | 0.20 | 0.39 | 0.21 | 0.11 | 0.07 | 0.14 | 0.34 | 0.25 | 0.60 | 0.64 | 0.07 | 0.20 | 0.27 | 0.27 |
| PAFRAC | 0.24 | 0.31 | 0.34 | 0.35 | -0.04 | 0.05 | 0.05 | 0.03 | 0.12 | 0.39 | 0.55 | 0.65 | 0.22 | 0.35 | 0.43 | 0.44 |
| CIRCLE_AM | 0.15 | 0.23 | 0.32 | 0.46 | -0.05 | -0.04 | -0.01 | 0.04 | 0.25 | 0.55 | 0.74 | 0.84 | 0.29 | 0.32 | 0.34 | 0.33 |
| CONTIG_SD | 0.17 | 0.32 | 0.31 | 0.33 | 0.15 | 0.33 | 0.34 | 0.31 | 0.38 | 0.70 | 0.79 | 0.81 | 0.12 | 0.01 | -0.07 | -0.12 |
| CONTIG_MD | 0.30 | 0.25 | 0.28 | 0.38 | 0.23 | 0.17 | 0.08 | 0.10 | 0.42 | 0.48 | 0.59 | 0.67 | 0.13 | 0.23 | 0.31 | 0.35 |
| AREA_MN | 0.38 | 0.27 | 0.37 | 0.47 | 0.27 | 0.24 | 0.41 | 0.57 | 0.37 | 0.31 | 0.49 | 0.59 | 0.09 | 0.13 | 0.20 | 0.19 |
| PARA_CV | 0.22 | 0.36 | 0.36 | 0.40 | 0.23 | 0.36 | 0.34 | 0.32 | 0.46 | 0.72 | 0.77 | 0.79 | 0.14 | 0.07 | 0.02 | -0.00 |
| PARA_RA | 0.10 | 0.30 | 0.35 | 0.48 | 0.10 | 0.29 | 0.35 | 0.32 | 0.19 | 0.58 | 0.72 | 0.75 | 0.26 | 0.26 | 0.26 | 0.27 |
| CONTIG_MN | 0.35 | 0.32 | 0.35 | 0.49 | 0.27 | 0.27 | 0.20 | 0.20 | 0.50 | 0.58 | 0.66 | 0.72 | 0.13 | 0.24 | 0.33 | 0.34 |
| CONTIG_RA | 0.14 | 0.33 | 0.38 | 0.49 | 0.14 | 0.33 | 0.37 | 0.34 | 0.29 | 0.65 | 0.74 | 0.76 | 0.28 | 0.26 | 0.26 | 0.28 |
| FRAC_MN | 0.40 | 0.39 | 0.38 | 0.49 | 0.19 | 0.22 | 0.16 | 0.18 | 0.48 | 0.65 | 0.75 | 0.79 | 0.22 | 0.30 | 0.34 | 0.34 |
| AREA_SD | 0.41 | 0.45 | 0.45 | 0.47 | 0.34 | 0.44 | 0.50 | 0.57 | 0.49 | 0.51 | 0.52 | 0.56 | 0.17 | 0.13 | 0.15 | 0.13 |
| AREA_AM | 0.49 | 0.49 | 0.47 | 0.46 | 0.39 | 0.48 | 0.53 | 0.55 | 0.53 | 0.53 | 0.50 | 0.51 | 0.17 | 0.12 | 0.11 | 0.09 |
| AREA_CV | 0.26 | 0.46 | 0.56 | 0.66 | 0.14 | 0.30 | 0.38 | 0.39 | 0.41 | 0.68 | 0.75 | 0.75 | 0.29 | 0.24 | 0.20 | 0.16 |
| AI | 0.34 | 0.42 | 0.46 | 0.54 | 0.32 | 0.40 | 0.41 | 0.39 | 0.64 | 0.80 | 0.85 | 0.89 | -0.03 | 0.06 | 0.14 | 0.16 |
| FRAC_CV | 0.30 | 0.47 | 0.48 | 0.56 | 0.11 | 0.32 | 0.36 | 0.35 | 0.39 | 0.71 | 0.82 | 0.86 | 0.31 | 0.32 | 0.34 | 0.33 |
| FRAC_SD | 0.31 | 0.48 | 0.48 | 0.56 | 0.12 | 0.32 | 0.35 | 0.35 | 0.40 | 0.72 | 0.82 | 0.86 | 0.31 | 0.32 | 0.34 | 0.34 |
| AREA_RA | 0.43 | 0.52 | 0.52 | 0.53 | 0.36 | 0.51 | 0.57 | 0.60 | 0.53 | 0.60 | 0.59 | 0.60 | 0.22 | 0.18 | 0.17 | 0.15 |
| FRAC_RA | 0.27 | 0.45 | 0.52 | 0.63 | 0.10 | 0.29 | 0.37 | 0.40 | 0.32 | 0.66 | 0.78 | 0.81 | 0.41 | 0.40 | 0.38 | 0.34 |
| LPI | 0.51 | 0.52 | 0.52 | 0.53 | 0.41 | 0.51 | 0.57 | 0.60 | 0.56 | 0.60 | 0.59 | 0.60 | 0.22 | 0.18 | 0.17 | 0.15 |
| GYRATE_MN | 0.43 | 0.39 | 0.51 | 0.59 | 0.29 | 0.32 | 0.46 | 0.50 | 0.50 | 0.60 | 0.79 | 0.83 | 0.17 | 0.27 | 0.32 | 0.30 |
| COHESION | 0.39 | 0.45 | 0.50 | 0.61 | 0.27 | 0.36 | 0.36 | 0.35 | 0.63 | 0.81 | 0.87 | 0.89 | 0.16 | 0.22 | 0.25 | 0.24 |
| SHAPE_MN | 0.44 | 0.49 | 0.54 | 0.61 | 0.22 | 0.34 | 0.43 | 0.45 | 0.47 | 0.68 | 0.81 | 0.85 | 0.19 | 0.28 | 0.32 | 0.31 |
| SHAPE_AM | 0.48 | 0.56 | 0.60 | 0.67 | 0.23 | 0.40 | 0.51 | 0.59 | 0.51 | 0.66 | 0.71 | 0.74 | 0.26 | 0.22 | 0.22 | 0.18 |
| GYRATE_CV | 0.30 | 0.50 | 0.57 | 0.63 | 0.17 | 0.40 | 0.51 | 0.56 | 0.43 | 0.74 | 0.84 | 0.87 | 0.31 | 0.27 | 0.26 | 0.23 |
| SHAPE_RA | 0.38 | 0.53 | 0.59 | 0.68 | 0.17 | 0.37 | 0.47 | 0.55 | 0.43 | 0.67 | 0.75 | 0.78 | 0.35 | 0.32 | 0.31 | 0.28 |
| GYRATE_AM | 0.52 | 0.55 | 0.56 | 0.59 | 0.39 | 0.50 | 0.55 | 0.58 | 0.61 | 0.70 | 0.71 | 0.71 | 0.24 | 0.20 | 0.20 | 0.16 |
| PLADJ | 0.45 | 0.48 | 0.50 | 0.57 | 0.40 | 0.44 | 0.43 | 0.41 | 0.73 | 0.84 | 0.88 | 0.90 | 0.20 | 0.20 | 0.21 | 0.20 |
| GYRATE_SD | 0.42 | 0.53 | 0.55 | 0.61 | 0.31 | 0.48 | 0.55 | 0.59 | 0.56 | 0.71 | 0.77 | 0.80 | 0.26 | 0.25 | 0.27 | 0.25 |
| CONTIG_AM | 0.46 | 0.48 | 0.50 | 0.58 | 0.41 | 0.45 | 0.44 | 0.42 | 0.73 | 0.84 | 0.88 | 0.90 | 0.20 | 0.21 | 0.21 | 0.19 |
| SHAPE_SD | 0.41 | 0.57 | 0.62 | 0.70 | 0.19 | 0.41 | 0.54 | 0.63 | 0.47 | 0.71 | 0.80 | 0.84 | 0.28 | 0.27 | 0.29 | 0.27 |
| SHAPE_CV | 0.37 | 0.56 | 0.62 | 0.71 | 0.18 | 0.42 | 0.55 | 0.63 | 0.46 | 0.73 | 0.83 | 0.87 | 0.31 | 0.28 | 0.29 | 0.27 |
| FRAC_AM | 0.47 | 0.56 | 0.62 | 0.72 | 0.23 | 0.40 | 0.49 | 0.55 | 0.55 | 0.76 | 0.86 | 0.90 | 0.29 | 0.28 | 0.28 | 0.25 |
| GYRATE_RA | 0.42 | 0.56 | 0.59 | 0.65 | 0.32 | 0.50 | 0.56 | 0.59 | 0.55 | 0.74 | 0.78 | 0.80 | 0.33 | 0.31 | 0.29 | 0.25 |
| ED | 0.42 | 0.49 | 0.53 | 0.58 | 0.29 | 0.42 | 0.49 | 0.52 | 0.40 | 0.60 | 0.70 | 0.71 | 0.55 | 0.56 | 0.57 | 0.51 |
| LSI | 0.42 | 0.49 | 0.53 | 0.58 | 0.29 | 0.42 | 0.49 | 0.52 | 0.40 | 0.60 | 0.70 | 0.71 | 0.55 | 0.56 | 0.57 | 0.51 |

**Figure A2. Pearson's correlation coefficients between all 51 landscape metrics and accuracy components.**

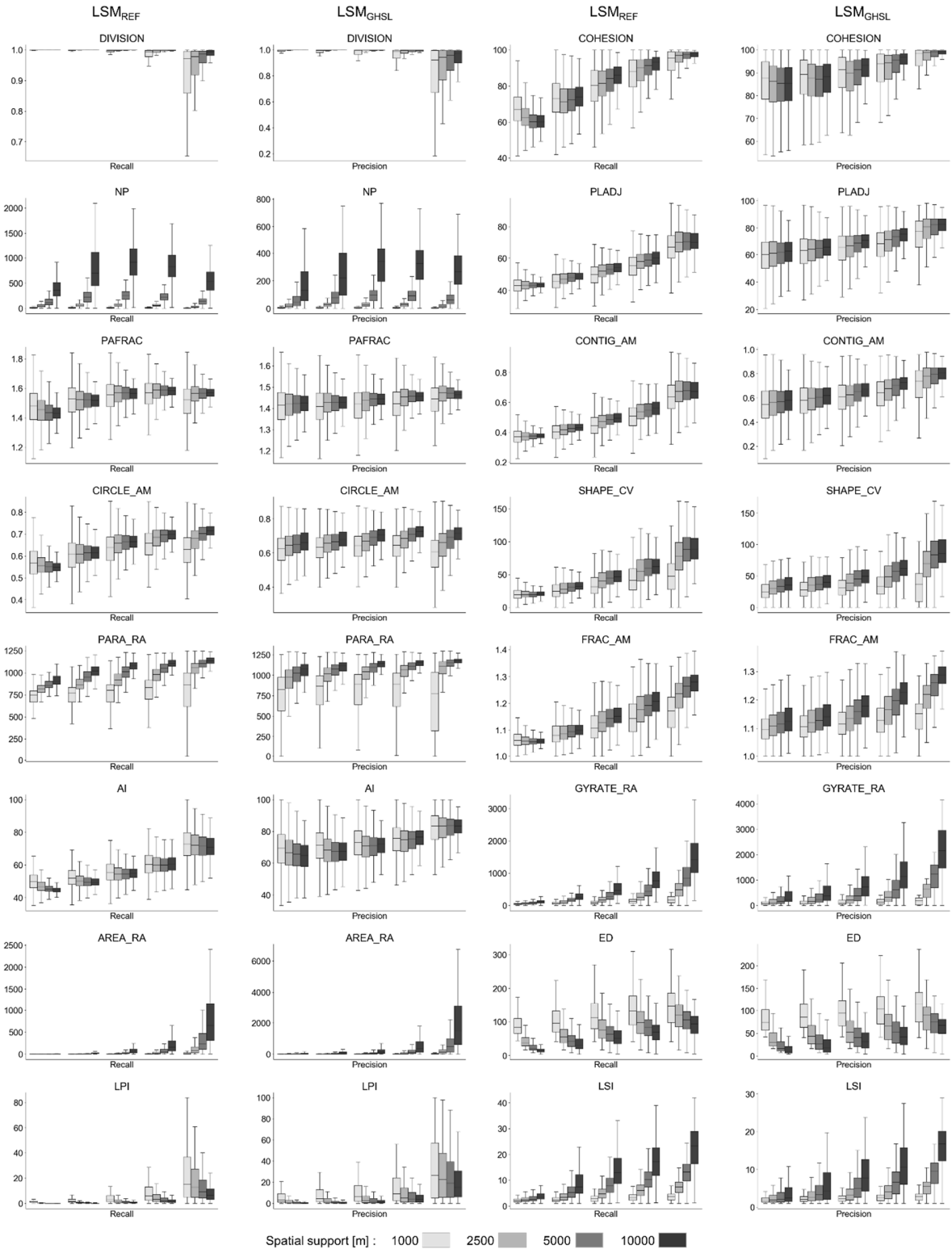

**Figure A3. Distributions of landscape metrics derived from the reference data (LSM$_{REF}$), and from the GHSL (LSM$_{GHS}$), within strata defined by quintiles of the response variables recall and precision, respectively. LSMs in the upper left exhibit least, in the lower right highest average correlation to the response variable across all support levels.**

**Figure A4. Sensitivity analysis: Correlation coefficients between the 51 landscape metrics and the accuracy estimates, over time and for different analytical units.**

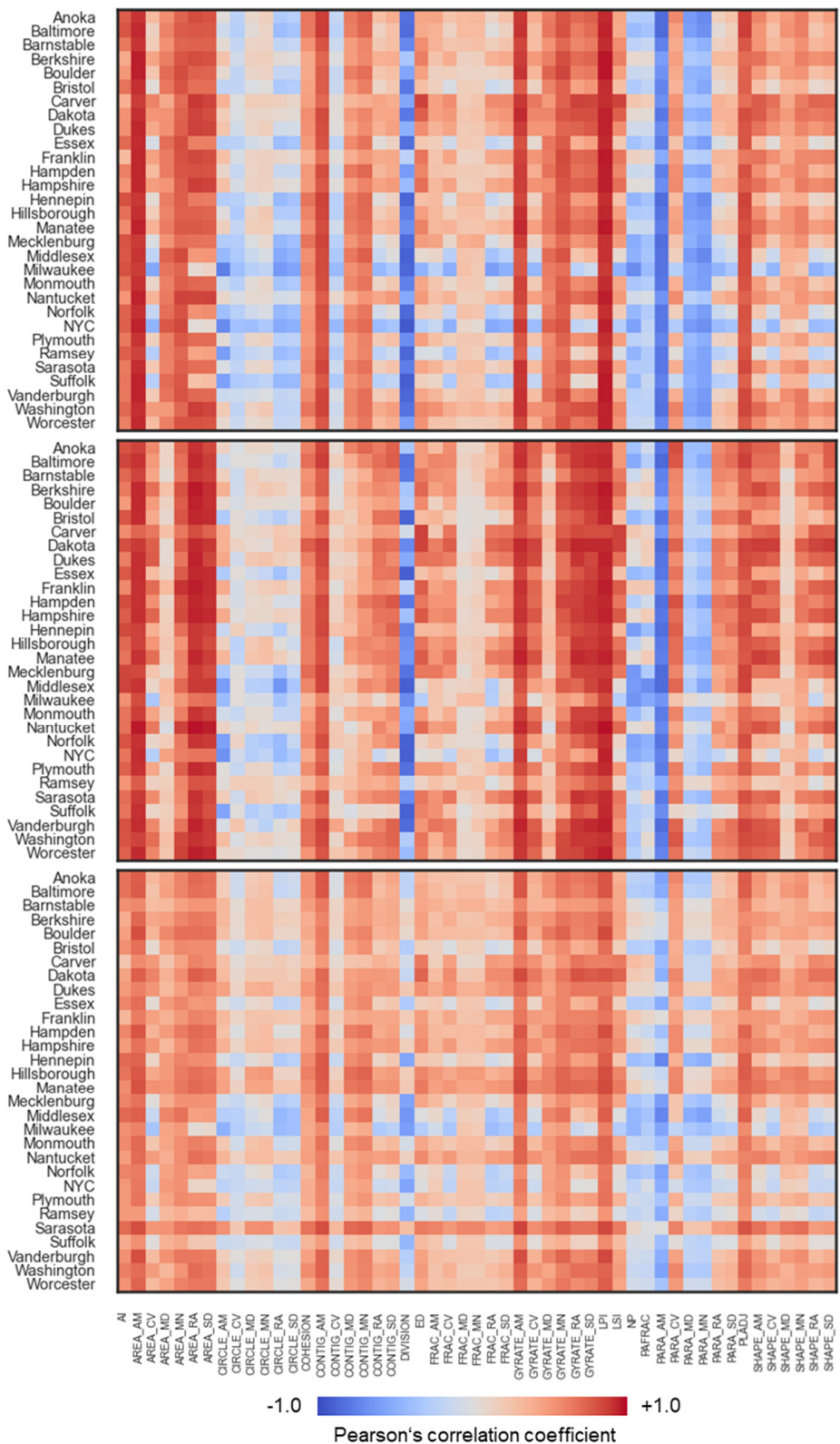

**Figure A5. Correlation of the 51 landscape metrics and built-up surface density in 30 U.S. counties, for 1km (top), 2.5km (middle), and 5km spatial support (bottom).**

23